\documentclass[aps,preprint,superscriptaddress]{revtex4}
\usepackage{amsmath}
\usepackage{graphicx}
\usepackage[english]{babel}

\begin{document}

\title{Finite element simulation of the liquid-liquid transition to metallic hydrogen}

\author{Matthew Houtput}
\affiliation{Theory of Quantum and Complex Systems, Universiteit Antwerpen, B-2000 Antwerpen, Belgium}

\author{Jacques Tempere}
\affiliation{Theory of Quantum and Complex Systems, Universiteit Antwerpen, B-2000 Antwerpen, Belgium}

\author{Isaac F. Silvera}
\affiliation{Lyman Laboratory of Physics, Harvard University, Cambridge, MA 02138}

\begin{abstract}
Hydrogen at high temperature and pressure undergoes a phase transition from a 
liquid molecular phase to a conductive atomic state, or liquid metallic hydrogen, sometimes referred to as the plasma phase transition (PPT).
The PPT phase line was observed in a recent experiment studying laser-pulse heated hydrogen in a diamond anvil cell in the pressure range $\sim 100-170$ GPa for temperatures up to $\sim 2000$ K. The experimental signatures of the transition
are (i) a negative pressure-temperature slope, (ii) a plateau in the heating curve, assumed to be related to the latent heat of transformation, and (iii) an abrupt increase in the reflectance of the sample.
We present a finite element simulation that accurately takes into account
the position and time dependence of the heat deposited by the laser pulse.
We calculate the heating curves and the sample reflectance and transmittance. This simulation confirms that the observed plateaus are related to the phase transition, however we find that large values of latent heat are needed and may indicate that dynamics at the transition are more complex than considered in current models. Finally, experiments are proposed that can distinguish between a change in optical properties due to a transition to a metallic state or due to closure of the bandgap in molecular hydrogen.

\end{abstract}
\maketitle

\section{Introduction}

\begin{figure}
    \centering
    \includegraphics[scale=0.6]{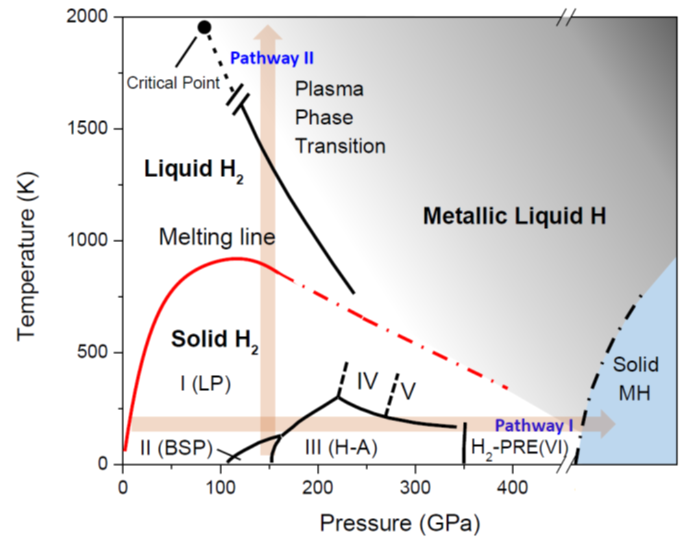}
    \caption{Recent phase diagram of hydrogen, combining experimental and theoretical results. It shows the two possible pathways for creating MH: I) the low-temperature pathway considered by Wigner and Huntington, and II) the high-temperature pathway to LMH. Figure extracted from \cite{Dias2017}.}
    \label{HPhaseDiagram}
\end{figure}

Metallic Hydrogen (MH) was studied by Wigner and 
Huntington \cite{Wigner1935} in 1935; they predicted that compressing solid molecular hydrogen 
(H$_2$) to a pressure (P) of 25 GPa or higher would destabilize the molecular bond and cause a phase transition from an insulating 
solid molecular state to a solid atomic metallic state. In its metallic form, hydrogen is expected to
have many interesting properties, amongst which high-temperature 
superconductivity stands out \cite{Ashcroft1968}.
After Wigner and Huntington's prediction it became clear that the
transition pressure lies at much higher pressure \cite{McMahon2012,McMinis2015}. The Wigner-Huntington transition has recently been observed at 495 GPa \cite{Dias2017}. The subject of this paper is a second pathway (Pathway II) to metallic hydrogen shown in Fig.~\ref{HPhaseDiagram}. This transition is at lower pressures and is achieved at high temperatures due to thermal dissociation to form liquid atomic metallic hydrogen (LMH).

The great outer planets such as Jupiter are mainly composed of hydrogen. Descending through Jupiter's atmosphere, the pressure and temperature (T) increase until a transition takes place to liquid metallic hydrogen. For planetary modeling the location of the transition temperature and pressure is very important \cite{Guillot2005}. An early theory based on a chemical model predicted high temperatures for the transition \cite{Saumon1992}. Density functional theories predicted much lower temperatures, and pressures in the 100-200 GPa range for the phase transition line \cite{Scandolo2003, Lorenzen2010, Morales2010}. The common denominator of these theories is a first-order phase transition from liquid molecular to liquid atomic metallic hydrogen with latent heat of transformation, a phase line with a negative P,T slope with increasing pressure, and a critical point (Fig.~\ref{HPhaseDiagram}), such that the transition is continuous for lower pressures and higher temperatures. However, DFT-based predictions had a large scatter of values because the results depend strongly on the choice of the density functional. An alternative route using quantum Monte-Carlo simulation has also been giving converging results recently \cite{Pierleoni2016,Mazzola2018}.

	The ideal experiment would have a thermally isolated sample in a pressure cell and add heat to the sample to warm it through the transition, enabling a determination of the transition P and T and the latent heat. In this case a plot of the temperature vs. heat added would have a rising slope with a plateau at the temperature of the transition to enable a direct measurement of the latent heat. There are two ways of achieving the high pressures and temperatures required to observe the PPT: dynamic (shock) compression and static compression in a diamond anvil cell (DAC). Neither of these techniques provide thermal isolation of the sample. In dynamic experiments P and T conditions last for tens of nanoseconds; in DACs high pressures can be maintained continuously. DACs cannot maintain their integrity if continuously heated to the high temperatures of the transition ($ \sim $1000-2000 K), so pulsed heating is utilized to heat the sample.
	
	Liquid metallic hydrogen was first observed in dynamic experiments using reverberating shock waves \cite{Weir1996}, however the measurement was not sensitive to determining the phase line. In a single-shock Hugoniot such as in Ref. \cite{Loubeyre2012} the temperature rise can be $\sim$10000-20000 K. A recent dynamic experiment on deuterium at the National Ignition Facility \cite{Celliers2018} investigated the transition at temperatures under 2000K. These dynamic experiments either did not probe the phase transition or did not determine the transition temperature.  In DACs one can measure both pressure and temperature (by fitting blackbody radiation to a Planck curve).  By using pulsed lasers to heat the sample for short periods of time (few hundred nanoseconds) a standard DAC can be used in which mainly the sample is heated and not the body of the DAC itself, enabling high static pressures and temperatures of thousands of degrees K.
	
	The technique of pulsed laser heating led to the recent experimental determination of the phase line of the PPT for pressures in the range of 100-170 GPa and temperatures up to ~2000 K by Zaghoo, Salamat, and Silvera (ZSS) \cite{Zaghoo2016}; these studies were further refined to determine the conductivity, degree of dissociation, and Drude parameters that lead to the dielectric function \cite{Zaghoo2017}. In the present paper we focus on this pathway to LMH. We note that the DAC is not a thermally isolated system and in the geometry used in the experiment, the energy from the pulsed laser heated sample can both go into heating the sample as well as diffuse into the diamonds (Fig.~\ref{ExpSetup}a).  Nevertheless a plateau was found in the heating curves; below we shall show that this can be simulated. The metallization transition changes the optical properties as the sample becomes absorbing and reflective. Detecting this change and fitting it to metallic optical conductivity has been the method used to detect the transition to the metallic phase. In particular, in \cite{Zaghoo2016}, the PPT is seen to lead to an abrupt increase in the reflectance, with frequency dependence in qualitative agreement with conductivity predicted by a Drude free-electron model of a metal. Additionally, the heating curve relating temperature to applied laser power is accompanied by the appearance of a plateau, indicating the presence of latent heat of transition.
  	Several authors \cite{Geballe2012,Knudson2015,Goncharov2017,Celliers2018} interpreted the plateau differently, instead relating it to a continuous change in optical parameters due to temperature dependent bandgap closure.
	
	In the experiment the properties change rapidly in time. The pressurized molecular hydrogen in the DAC is heated by a pulsed laser beam (280 ns pulse width), partially absorbed by a thin tungsten layer deposited on one of the diamonds (Fig.~\ref{ExpSetup}). The heated tungsten film heats the hydrogen pressed to its surface, creating a layer of LMH when sufficiently hot. This layer appears, thickens, and then shrinks as the pulse cycle is traversed. Increased laser power leads to thicker films of LMH growing at the LMH-molecular H$_2$ interface. Continuous wave laser light shines on the sample to measure transmission and reflection during the heating cycle using time-resolved spectroscopy.  The optical properties are strongly time-dependent and moreover feed back into the position dependent absorption of the heating laser pulse, as the LMH itself also absorbs the laser power.
	
	In this paper we simulate these dynamics with a finite element analysis (FEA) of the pulsed-laser heated DAC.  Previous simulations \cite{Geballe2012, Montoya2012} investigated the case of a thicker ($\sim \mu$m) laser absorber embedded in an insulating material, with double sided heating; the absorber itself developing a steep thermal gradient.  In these cases the absorber was also the sample.  In the experiment of ZSS the absorber is only several nanometers thick, heats uniformly and heats the sample. A recent FEA by Goncharov and Geballe \cite{Goncharov2017} attempting to simulate the experiment of ZSS did not find a plateau.  Their simulation did not include thermodynamic properties of LMH, only molecular hydrogen so that it was not designed to simulate the experiment and reported non-physical results \cite{Silvera2017}.

The goal of the present paper is to provide a finite element simulation that better 
corresponds to the experimental configuration of ZSS \cite{Zaghoo2016}.
We show that it is important to 
take into account the spatial and temporal dependence of the heat delivered to the
cell by the laser pulse. When this improvement is taken into account, we find that 
plateaus can appear in the heating curve, due to the latent heat of transformation.
The temperature at which the heating curve develops the plateau is essentially the 
metallization temperature and can be used as an estimate for this temperature.

The structure of this paper is as follows.
In Sec.\ref{Methods}, we outline the numerical simulation method, consisting
of three computation steps per time step: the laser-pulse heating, 
the heat conduction, and the temperature change.
The results are presented in Sec.\ref{Results} for the
heating curves and 
for the reflectance and transmittance.
In Sec.\ref{Discussion} these results are discussed and compared to other simulations and
to the experiment. The importance for further experiments is discussed in Sec.\ref{PhaseLine}. We conclude in Sec.\ref{Conclusion}.

\section{Numerical methods \label{Methods}}

\begin{figure}[t]
\centering
\includegraphics[scale=0.17]{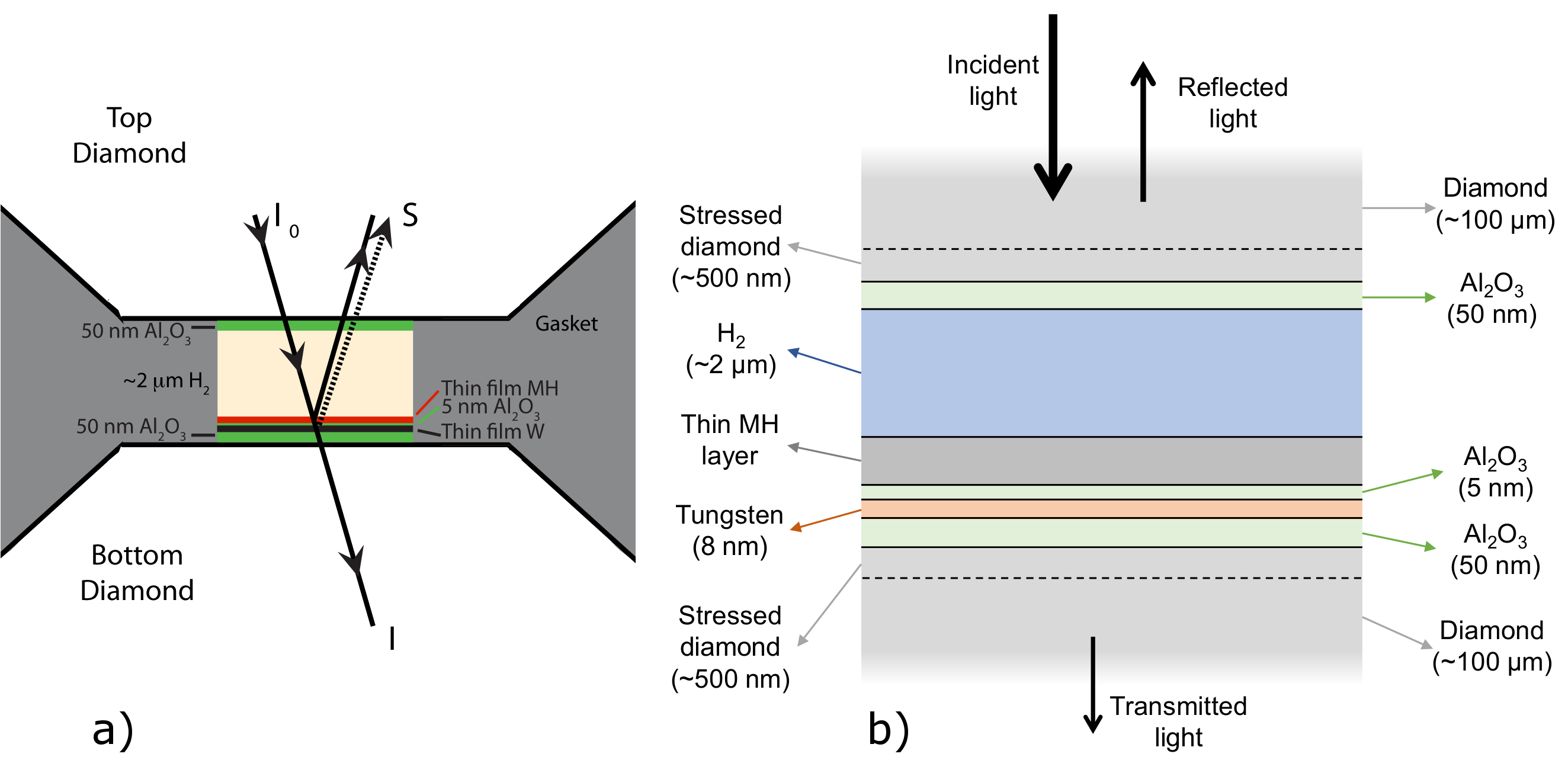}
\caption{\label{ExpSetup} a) The heart of the diamond anvil cell used by ZSS (not to scale), including the gasket and the diamond culets. Al$_2$O$_3$ signifies amorphous alumina. The heating laser beam is not shown, only light beams for measuring transmittance and reflectance. b) A one-dimensional model of the sample chamber used in the simulation. The diamond-air interfaces at the top and bottom are taken into account but are not shown. Stressed diamond in the culet region will be deformed and have additional defects, such as vacancies, and thus will have a lower thermal conductivity. In both panels we take the $z$-direction to be perpendicular to the surfaces.}
\end{figure}

The experimental setup of Ref.~\cite{Zaghoo2016} is shown in Fig.~\ref{ExpSetup}a.
The sample chamber is about 20 $\mu$m in diameter and has cylindrical 
symmetry. It consists of multiple thin layers, as modeled and illustrated schematically
in Fig.~\ref{ExpSetup}b. The largest portion of the sample chamber consists 
of a 2 $\mu$m thick layer of molecular hydrogen, transparent to the heating laser.
The diamond culets are covered with a 50 nm layer of amorphous alumina (Al$_2$O$_3$); an $\sim 8$ nm layer of tungsten (W) is deposited on one of the diamonds, which is covered by a 5 nm Al$_2$O$_3$ layer. The transparent alumina coatings inhibit the diffusion of hydrogen into the diamonds and the tungsten. The partially transmitting tungsten layer acts as an absorber of laser light.  A pulsed heating laser (with pulse duration $\tau = 280$ ns and wavelength $\lambda = 1064$ nm) is incident on the top (hydrogen side) to heat the sample.  Upon heating to the metallization temperature, part of this molecular hydrogen layer converts into metallic hydrogen.

Probe laser light (wavelengths of  514 nm, 633 nm, 808 nm, and 980 nm)
is continuously shone into the cell from the top (hydrogen side) to measure the transmittance and reflectance.
Both the heating and probe lasers are slightly defocused, so 
that the light intensity can be assumed roughly uniform along the surface of the cell.
Finally, the temperature is determined by collecting the blackbody radiation from the top 
of the DAC and fitting it to a Planck curve to enable determination of the peak temperature in a pulse \cite{Rekhi2003}.

In our numerical simulation, we use a one-dimensional (1D) model that
incorporates the different layers with dimensions shown in Fig.~\ref{ExpSetup}b.
We also include 100 $\mu$m of the bottom diamond in our simulation and assumed,
as our boundary conditions, that the rest of the (2 mm thick) diamond, as well as
the top diamond, remains at room temperature during the heating pulse.
Since our model is 1D, neither the conductive energy loss due to the gasket nor the Gaussian profile of the heating laser can be taken into account. However, both the sample chamber and the laser profile are wide in the radial direction, meaning the 1D model acts as a good first approximation able to capture the essential physics.

To describe the thermodynamics of the system, the one-dimensional form of the diffusion equation was used. We define the z-axis to be normal to the plane of the sample. Call $q(z,t)$ the local energy density ($J/m^3$), $T(z,t)$ the temperature, $S(z,t)$ the Poynting vector of the heating laser in the positive z-direction, and $k$ the thermal conductivity of the material. Then, the continuity equation implies:
\begin{equation} \label{QconEq}
    \frac{\partial q}{\partial t} = -\frac{\partial S}{\partial z} + \frac{\partial}{\partial z} \left( k \frac{\partial T}{\partial z} \right).
\end{equation}
To proceed, this equation was discretized in the z- and t-domains. A uniform temporal grid $t_n = n \Delta t$ and a non-uniform spatial grid $z_i$ with element thicknesses $\Delta z_i$ were chosen. Replacing the energy density per unit volume $q(z,t)$ by the energy density per unit surface $q_i^{(n)} = q(z_i,t_n)\Delta z_i$, and defining $\Delta q_i^{(n)} := q_i^{(n+1)}-q_i^{(n)}$, equation (\ref{QconEq}) becomes:
\begin{equation} \label{DeltaQ}
    \Delta q_i^{(n)} = \Delta t \left[ S^{(n)}\left(z_i - \frac{\Delta z_i}{2} \right)-S^{(n)}\left(z_i + \frac{\Delta z_i}{2} \right) + \frac{T^{(n)}_{i+1}-T^{(n)}_i}{r^{(n)}_{i,i+1}} - \frac{T^{(n)}_i-T^{(n)}_{i-1}}{r^{(n)}_{i-1,i}} \right].
\end{equation}
Here a superscript $^{(n)}$ was used to indicate a variable is evaluated at time $t_n$, and $r^{(n)}_{i,i+1}$ denotes the thermal resistance between elements $i$ and $i+1$, given by:
\begin{equation} \label{ResistanceDef}
r^{(n)}_{i,i+1} = \frac{z_{i+1}-z_i}{k^{(n)}_{i,i+1}},
\end{equation}
with $k^{(n)}_{i,i+1}$ the thermal conductivity of the material between elements $i$ and $i+1$.

 As an amount of heat $\Delta q_i^{(n)}$ is added to the element $i$,
 one of two things can happen. First, the temperature in the element can increase.
 Second, if element $i$ is a hydrogen element, the hydrogen molecules can
 dissociate, creating metallic hydrogen. To model this, a metallization
 fraction $f_i^{(n)}$ between $0$ and $1$ is assigned to the hydrogen elements:
 $f_i^{(n)} = 0$ means the element is in the molecular state; when $f_i^{(n)} = 1$ it is metallic.

For our grid, we chose elements of thickness $\Delta z_i = 1$nm in the hydrogen and alumina layers, and a non-uniform grid of thicknesses $\Delta z_i > 5$nm in the lower diamond. The tungsten is considered to be its own element with uniform temperature. Both doubling and halving the number of elements did not change the results.

We start with a uniform temperature of $T_j^{(0)}=300$K and no MH: $f_j^{(0)} = 0$. Every time step in the simulation is divided in three parts.
First, the Poynting vector $S^{(n)}(z)$ is determined in every element from the optical properties of each element. Second, the thermal conductivity between adjacent elements and the contribution of heat conduction to $\Delta q_i^{(n)}$ is calculated.
Third, $\Delta q_i^{(n)}$ is calculated from (\ref{DeltaQ}) and is used to compute the resulting new temperature and metallization fraction for each element. The simulation proceeds in time by repeating this process with the new values $T_i^{(n+1)}$ and $f_i^{(n+1)}$, which allows us to calculate $\Delta q_i^{(n+1)}$, and so on. In the following subsections, each of these parts is described in detail. The values for the material parameters are listed in the final subsection.

\subsection{Heat absorption from the laser pulse: calculation of the Poynting vector \label{LaserHeat}}
The Fresnel formalism is suitable to describe the optical properties of 
any number $N$ of thin layers (thinner than the wavelength $\lambda$). 
The layers in the chosen geometry are shown in Fig.~\ref{ExpSetup}b. 
The complex index of refraction for layer $j$ is denoted by 
$\tilde{N}_j = n_j + \textrm{i} K_j$, and 
$\kappa_j = \frac{2 \pi}{\lambda} \tilde{N}_j$ is
the corresponding complex wavenumber. Furthermore, $d_j$ are the thicknesses of the layers, and $Z_j$ 
are the positions of the boundaries between the layers.
The thickness of the layer of
metallic hydrogen is found from the metallization
fractions $f_i^{(n)}$ mentioned above. In our simulation, we always
find a sharply bounded region where $f_i^{(n)} = 1$, so the 
layer thickness of metallic hydrogen can be clearly determined.
For a mixture of molecular and metallic hydrogen, the following complex index of refraction was used \cite{LandauLifshitz}:
\begin{equation}
\tilde{N}_{\text{eff}}^{\frac{2}{3}} = f \tilde{N}_{\text{MH}}^{\frac{2}{3}} + (1-f) \tilde{N}_{\text{H}_2}^{\frac{2}{3}}.
\end{equation}

The laser is assumed to be incident along the normal, so
the electric field is tangential to the interfaces between the layers.
The magnitude of the electric field throughout the sample is given by the piecewise
function
\begin{eqnarray}
E(z) & = & E_j(z) \text{ where } Z_j \leq z \leq Z_{j+1}, \\
E_j(z) & = & a_j e^{i \kappa_j (z-Z_j)} + b_j e^{-i \kappa_j (z-Z_j)} \label{Efield},
\end{eqnarray}
with unknown coefficients $a_j$ and $b_j$. The boundary conditions are such that $E$ and 
$\partial E / \partial z$ are continuous across the boundaries. 
Applied to (\ref{Efield}), this gives the following $2N-2$ equations:
\begin{equation}
 \label{Fresneleqn}
\left\{
\begin{array}{rcl}
a_j e^{i \kappa_jd_j} + b_j e^{-i\kappa_jd_j} & = & a_{j+1} + b_{j+1} \\
a_j \kappa_j e^{i\kappa_jd_j} - b_j \kappa_j e^{-i\kappa_jd_j} & = & a_{j+1} \kappa_{j+1} + b_{j+1} \kappa_{j+1}
\end{array}
\right.
\hspace{5pt}
.
\end{equation}
Imposing these, along with the boundary conditions
\begin{equation}
\left\{
\begin{array}{l}
    a_1 = 1  \\
    b_N = 0
\end{array}
\right.
\hspace{10pt}
\text{or}
\hspace{10pt}
\left\{
\begin{array}{l}
    a_1 = 0  \\
    b_N = 1
\end{array}
\right.
\hspace{5pt}
,
\end{equation}
depending on the incident direction of the laser, gives a system with a unique solution for all $a_j$ and $b_j$, which is found numerically. Calculating the time-averaged Poynting vector and normalizing it to the time-dependent intensity of the laser pulse $I_0(t)$ then gives the laser power profile as a piecewise function in space:
\begin{equation}
S_j^{(n)}(z) = -I_0(t_n) \times \left[
\begin{array}{cc}
& \frac{n_j}{n_1} \left( |a_j|^2 e^{-\frac{4\pi}{\lambda}K_j(z-Z_j)}- |b_j|^2 e^{\frac{4\pi}{\lambda}K_j(z-Z_j)} \right) \nonumber \\
& - \frac{K_j}{n_1} 2 |b_j^* a_j| \sin \left(\frac{4 \pi}{\lambda} n_j (z-Z_j) + \text{arg}(b_j^* a_j) \right) \\
\end{array}
\right]. \label{Poynting} 
\end{equation}
Note that $S_j(z)$ is constant if $K_j = 0$, meaning that no power is lost in non-absorbing 
materials as expected. The Poynting vector is a continuous function of $z$ because of the 
imposed electromagnetic boundary conditions. For the time-dependent laser pulse intensity we use the following function, that simulates the pulses used in the experiment:
\begin{equation} \label{pulseFunction}
I_0(t) = \frac{2 A t}{\tau^2} e^{-\left(\frac{t}{\tau}\right)^2},
\end{equation}
where $A$ is the pulse energy (per unit area) and $\tau$ the full width at half maximum (FWHM)
time. The transmittance $Tr$ and reflectance $R$ of this multilayered 
system are calculated by comparing the transmitted and reflected laser power to the 
incident laser power assuming the outermost layers are not absorbing ($K_1=K_N=0$). We get
\begin{equation} \label{TransRef}
 Tr = \frac{n_N}{n_1}|a_N|^2, \ \ \ \ R = |b_1|^2.
\end{equation}
The transmission and reflection are calculated at every point in time for 
a wavelength of 500 nm.

For thickness variations comparable to the 
wavelength of light, the Fresnel equations lead to Fabry-P\'erot type
interferences. However, this is an artifact related to the assumption of 
coherent, collimated illumination. Since the heating laser is not collimated,
the cell chamber should not be treated as a Fabry-P\'erot interferometer.
To correct for this, the sample chamber is divided into the following subsystems:
\begin{enumerate}
\item The air-diamond interface at the top
\item The alumina coated boundary between the top diamond and the hydrogen
\item The ``active region'', starting from the H$_2$-MH interface down to the 
interface between alumina and the lower diamond, and
\item The diamond-air interface at the bottom.
\end{enumerate}

Absorption takes place only in the active region. This region is treated within
the Fresnel formalism. The remaining boundaries are treated in the incoherent approximation: 
only the \textit{intensity} changes from reflection and transmission are taken 
into account. 
The reflection coefficients used for the boundaries outside
the active region are fixed, and are listed in table \ref{ReflCoefs}. The (multiple) reflections between the boundaries outside the
active region lead to power influx on the active region from both sides. Using the previously outlined Fresnel method on the active region with this power influx as input, the heat deposited in each element can be calculated.

\begin{table}
\begin{tabular}{l||l|l|l|}
Boundaries & $R_1$ & $R_2$ & $R_4$ \\ \hline \hline
 $\lambda = 500 $nm & 0.174 & 0.011 & 0.174 \\ \hline
 $\lambda = 1064 $nm & 0.168 & 0.003 & 0.168 \\ \hline
\end{tabular}
\caption{\label{ReflCoefs} Reflectance coefficients $R$ of the non-active boundaries. 
Transmittance can be found using $Tr = 1-R$. The indices refer to the interfaces as
listed in the text.}
\end{table}

\subsection{Heat redistribution from thermal diffusion: calculation of the thermal resistance \label{DiffHeat}}

If elements $i$ and $i+1$ are associated with the same material, the thermal resistance can be calculated with formula (\ref{ResistanceDef}) where $k_{i,i+1}:=k$ is the constant thermal conductivity of that material. For the hydrogen elements, we need to be more careful, since in general, the thermal conductivity of hydrogen depends on its state. Therefore, the following interpolation is used \cite{Woodside1961}:
\begin{equation} \label{kHydrogen}
k_{i,i+1} = k_{\text{H}_2}^{1-<f>}k_{\text{MH}}^{<f>} \ , \ <f>=\frac{f_i + f_{i+1}}{2}.
\end{equation}
The above method works for all pairs of neighboring elements except those where either $i$ or $i+1$ is the tungsten element.
The thermal resistance to the tungsten element from either side is taken as the thermal resistance of 
the respective alumina layer. This includes any thermal boundary resistance. In general, the thermal resistance of an alumina layer with thickness $d$ can be written as follows:
\begin{equation}
    r_{\text{Al}_2\text{O}_3} = \frac{d}{k_{\text{Al}_2\text{O}_3}} + r_{\text{bound},1} + r_{\text{bound},2}.
\end{equation}
It is possible that the thermal boundary resistances $r_{\text{bound},1}$ and $r_{\text{bound},2}$ cannot be neglected, since the thermal boundary resistance of amorphous alumina to platinum 
\cite{Cappella2013} and to aluminum \cite{Hopkins2007} is of the same order as the bulk thermal resistance 
of the alumina layers.

\subsection{Change in temperature and metallization fraction \label{TempChange}}
Let $\Delta q_i^{(n)}$ be the total heat generated in element $i$. This is the sum of the heat delivered
by the laser pulse described in subsection \ref{LaserHeat} and the heat conducted 
into the element as discussed in subsection \ref{DiffHeat}. From the mass density $\rho$ and the specific heat per unit mass $c$ of the material, the specific heat per unit volume $\rho c$ is calculated. From its definition, it follows that the change in temperature is given by:
\begin{equation} \label{Tchange}
\Delta T_i^{(n)} := T_i^{(n+1)}-T_i^{(n)} = \frac{\Delta q_i^{(n)}}{\rho c \Delta z_i}.
\end{equation}
For elements with a mixture of metallic and molecular hydrogen, 
the following relation based on the Dulong-Petit form is assumed:
\begin{equation} \label{rhocInterp}
\rho_{\text{eff}} c_{\text{eff}} = f \rho_{\text{MH}} c_{\text{MH}} + (1-f) \rho_{\text{H}_2} c_{\text{H}_2}.
\end{equation}
For the hydrogen elements the temperature can be increased up to the metallization temperature $T_m$.
In this paper we present a simulation at one pressure condition ($P=140$GPa) corresponding to a transition temperature of $T_m=1500 K$.
Then, we assume a first-order phase transition occurs, which can be modeled by 
changing the metallization fraction as follows:
\begin{equation} \label{fchange1}
\Delta f_i^{(n)} := f_i^{(n+1)}-f_i^{(n)} = \frac{\Delta q_i^{(n)}}{\rho L \Delta z_i},
\end{equation}
where $L$ is the latent heat per unit mass and $\rho L$ is the latent heat per unit volume. The metallization fraction can increase until it reaches $f_i = 1$, after which any additional heat again leads to a temperature increase. If $\Delta q_i^{(n)}$ is negative, the temperature/metallization fraction decreases, so the process occurs in reverse order.

\subsection{Values for the material parameters \label{SecMatParams}}

\begin{table*}[t]
\centering
\begin{tabular}{l||l|l|l|l|l|l|}
 & H$_2$ & MH & Tungsten & Al$_2$O$_3$ & Diamond & Stressed diamond \\ \hline \hline
$\rho$ $\left(\frac{kg}{m^3} \right)$ & 850 & 850 & 25100 & 5000 & 4400 & 4400 \\ \hline
$k$ $\left( \frac{W}{m K} \right)$ & 20 & 20 & - & 20 & 2200 & 220 \\ \hline
$\rho c$ $\left( \frac{MJ}{m^3 K} \right)$ & 10 & 20 & 3.5 &  1.2 & 1.7 & 1.7 \\ \hline
$\tilde{N}(\lambda = 1064 \text{nm})$ & 2.31 & 5.22+6.48i & 3.12+3.67i & 2.16 & 2.39 & 2.39 \\ \hline
$\tilde{N}(\lambda = 500 \text{nm})$ & 2.46 & 2.94+4.58i & 3.31+2.74i & 2.20 & 2.43 & 2.43 \\ \hline
$\rho L$ $\left( \frac{MJ}{m^3} \right)$ & \multicolumn{2}{|c|}{1700} & - & - & - & - \\ \hline
$T_m (K)$ & \multicolumn{2}{|c|}{1500} & - & - & - & - \\ \hline
Sources & \multicolumn{2}{|c|}{\cite{Rillo2019,Hanfland1994,Evans1998,Morales2010,Moroe2011}} & \cite{Haynes2015,Raju1997} & \cite{Malitson1963,Duan1999} & \cite{Haynes2015,Olson1993} & - \\ \hline
\end{tabular}
\caption{\label{MatParams} The values for the material parameters used in the simulation. A dash means the parameter is not used in the simulation. Stressed diamond is diamond with a tenfold lower thermal conductivity.}
\end{table*}

The material parameters used are given in Table~\ref{MatParams}. These values 
correspond to estimates at 140 GPa. For the thermal conductivities of molecular 
hydrogen and amorphous alumina no reliable values were found. However, from studies of similar materials in the literature we could make an estimate of these values. For H$_2$, an estimate is made using values measured for a pressurized gas of normal H$_2$ \cite{Moroe2011} at pressures of up to $\sim$100 MPa. Extrapolation of these results leads to a thermal conductivity of the same order of magnitude as that of MH. The specific heat 
and the optical properties were obtained by extrapolating values measured at ambient 
conditions to high pressure, using the Vinet equation of state \cite{Vinet1987}. 
The complex index of refraction was extrapolated using the Drude model for metals, 
and the Clausius-Mossotti relation (in both cases assuming that the 
dielectric function $\epsilon$ satisfies $\epsilon = \tilde{N}^2$). For $H_2$, the index of refraction measured at high pressure \cite{Evans1998} was extrapolated to the desired wavelengths using the Lorentz model.
The FWHM time duration of the laser pulse was chosen as $\tau$ = 280 ns. The laser pulse energy per 
unit area is varied in the range $47$ kJ/m$^2 \leq A \leq 54 \ $kJ/m$^2$. A thermal 
boundary resistance of $10^{-7}$ m$^2$ K/W was added between amorphous alumina 
and diamond. The other thermal boundary resistances were assumed small enough to be neglected.

\section{Simulation results \label{Results}}

\subsection{Heating curves \label{ResultsA}}
\begin{figure}[t]
 \centering
 \includegraphics[scale=0.55]{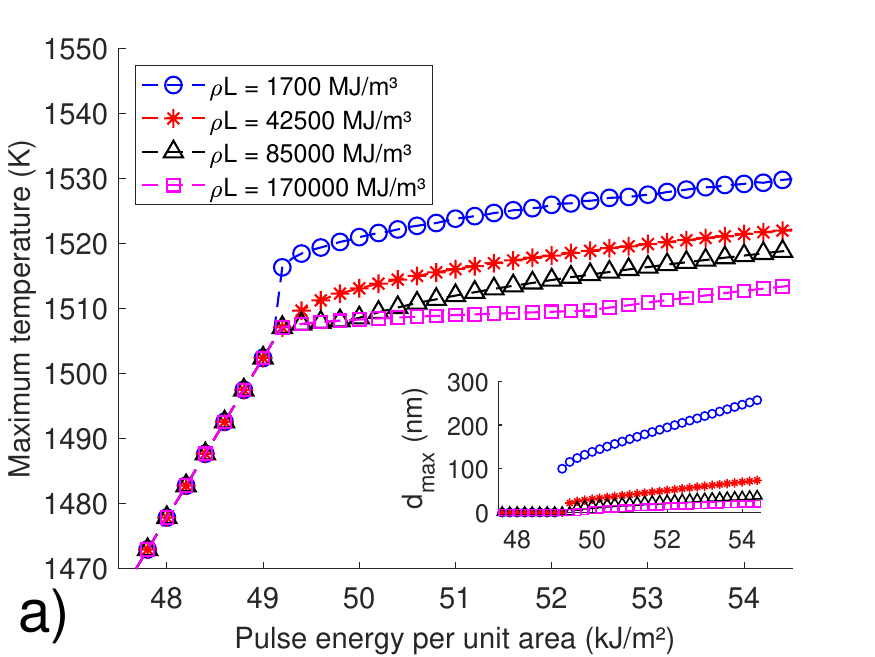}
\includegraphics[scale=0.25]{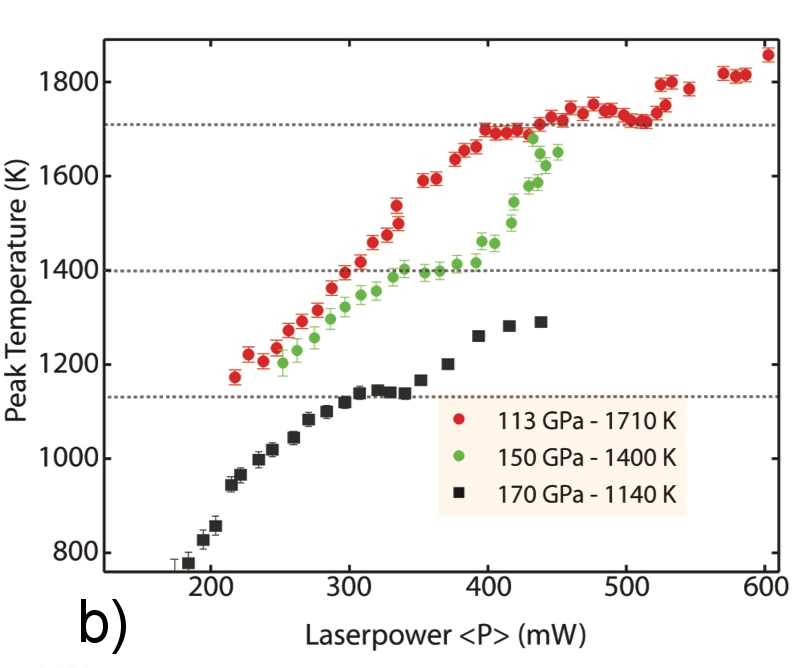}
 \caption{ \label{SimHeatingCurves} a) Simulated heating curves for the material parameters given in Table~\ref{MatParams}. Additionally, the latent heat was varied over a large range. The inset shows the maximum thickness of the MH layer. b) Heating curves as experimentally measured by ZSS
\cite{Zaghoo2016}. The plateaus vary in width and shape as the pressure varies, but it is clear 
that the slopes before and after the plateau have similar values, in contrast to the simulated heating curves.}
\end{figure}
In order to compare with the experimental results \cite{Zaghoo2016}, 
heating curves were simulated. These show the temperature in the sample as a function of the input pulse energy. Experimentally, the peak temperature in a heating cycle is determined from a fit of the blackbody radiation, using a procedure discussed by Rekhi et al. \cite{Rekhi2003}. Numerically, this temperature is approximated by the maximum temperature at any point in time in any element $T_{max}$. Such a procedure is necessary since no steady state is reached. The heating curves are found by plotting $T_{max}$ as a function of the pulse energy per unit area. The result is shown in Fig.~\ref{SimHeatingCurves}a); the heating curves measured by ZSS \cite{Zaghoo2016} are shown in Fig.~\ref{SimHeatingCurves}b).

In Fig.~\ref{SimHeatingCurves}a), the open circles connected by a (blue) dashed line show the results obtained with the
parameters from Table~\ref{MatParams}. For pulse energies above $49.2$ kJ/m$^2$ 
metallization is reached; the temperature scales linearly for lower pulse energy.
When the pulse energy is high enough to reach metallization, the absorption strongly increases,
leading to a quick growth of the layer of metallic hydrogen and an abrupt rise in temperature, disallowing a plateau.
For higher pulse energies, the absorption mainly takes place at the boundary between 
the metallized layer and the molecular hydrogen as the tungsten heater is increasingly
obscured by the MH layer. Additionally, the pulse energy goes into thickening of the layer
rather than to further increase the temperature. The object of a simulation is to reproduce the experimental observation. The absence of a plateau 
in the heating curve does not agree with the experimental observation \cite{Zaghoo2016}.

In order to obtain heating curves with a plateau, it proved
necessary to substantially increase the latent heat of transformation. This removes the rapid layer growth, and allows the tungsten layer to remain active as a heater.
By increasing the latent heat by factors of 25, 50 and 100, the curves with asterisks, triangles and squares (Fig.~\ref{SimHeatingCurves}a) were obtained, respectively.
The sudden jump in temperature disappears, and is gradually replaced by a plateau 
if the latent heat is larger than $85000$ MJ/m$^3$.
As the latent heat is increased, the width of the plateau increases and the 
thickness of the MH layer decreases.
It should be noted that the MH layer remains thin ($\leq 15$ nm) for any point on the plateau, 
so that both the tungsten and the MH layer absorb the laser energy and act as heating sources in this case.
The resulting slope of the heating curve after the plateau is significantly 
smaller than the slope before the plateau. At the onset of the transition
(at pulse energies of $49.2$ kJ/m$^2$), the maximum temperature in the DAC is 
located at the tungsten and is slightly higher than the transition temperature $T_m = 1500$ K,
as the thin (5 nm) alumina layer on the tungsten acts as a small thermal impedance.

\begin{figure}[t]
\centering
\includegraphics[scale=0.6]{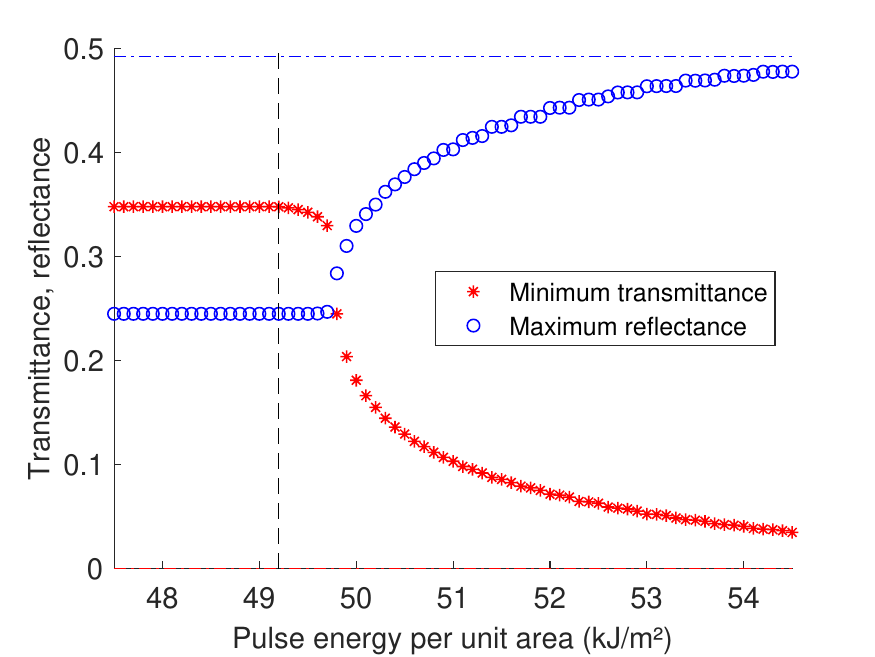}
\caption{ \label{TRcomp} Unnormalized minimum transmittance and maximum reflectance at $\lambda = 500$ nm reached during the simulation runs, as a function of pulse energy. We chose $\rho L = 170000$ MJ/m$^3$, that is, 100 times the expected value. The transition begins at the vertical dashed line, where transmittance and reflectance both change; below this energy the transmission and reflection are due to the tungsten film.}
\end{figure}

\subsection{Reflectance and transmittance \label{ResultsB}}

In the experiment \cite{Zaghoo2016}, the presence of the plateau is taken as a sign of a first-order phase transition.
On the other hand, several authors \cite{Celliers2018,Rillo2019} have suggested that the plateau corresponds to the onset of absorption rather than metallization. 
A recent Quantum Monte Carlo calculation \cite{Rillo2019} indicates that the absorption starts to increase near the experimental plateaus, whereas the increase in reflection occurs at higher temperatures/pressures. This highlights the difficulty of correctly identifying the transition pressure and temperature. Here, we will follow  
Ref.~\cite{Zaghoo2016} and interpret the onset of the plateau as indicative of the phase transition.

To determine whether the phase that results after the transition contains free electrons, ZSS measured the transmittance and reflectance and compared the results to a Drude free-electron model. It has also been pointed out that the free-electron model is not valid at the transition, but applies beyond the phase line when the electron density is higher \cite{Zaghoo2017, Morales2010}. In the plateau region close to the transition, molecules dissociate and reform on a short time scale, so the charge density exists but is small \cite{Morales2010}. Experimentally \cite{Zaghoo2016} the reflectance change is very small in this region, whereas the absorbance (transmittance change) is larger and grows with thickening of the MH. Density functional theory \cite{Morales2010} predicts a reasonable but small optical conductivity, yet a small but finite DC conductivity in the plateau region. The DC electrical conductivity seems to be a good order parameter and should have a discontinuity for this first order phase transition. The onset of the plateau correlates with the electron density and seems to be a proper choice for the transition line, not the abrupt rise of the reflectance that starts deeper in the plateau region. 

Using expression (\ref{TransRef}) the transmittance and reflectance can also be
calculated in our simulation. The minimal transmittance and maximal reflectance as 
a function of time are plotted as a function of input pulse energy, shown in 
Fig.~\ref{TRcomp} for $\lambda = 500$ nm and $\rho L = 170000$ MJ/m$^3$. Both the transmittance 
and reflectance remain constant before the transition starts, since there is no change in optical
parameters in our simulation. When the transition starts, the transmittance and reflectance change due to the layer of MH that is forming. As the pulse energy is increased, the MH layer grows thicker and 
thicker. The transmittance approaches zero and the reflectance approaches its 
bulk value. Near the final data points of Fig.~\ref{TRcomp}, which correspond to a MH 
layer thickness of $\sim 20$ nm, the transmittance and reflectance are almost converged to the bulk value.

\section{Discussion \label{Discussion}}
In this paper we have used the proper geometry and thermodynamic properties for the experimental demonstration of LMH \cite{Zaghoo2016, Zaghoo2017}, whereas earlier simulations are not relevant to this experiment \cite{Geballe2012, Montoya2012, Goncharov2017}.  We have found that the results of the simulation are quite sensitive to material properties such as thermal conductivity (TC) and latent heat, but also the optical properties of all materials in the DAC, as these play a large role in determining the absorbed laser power. We explored a large range of parameters. Of particular importance is the thermal conductivity of various materials and we have attempted to consider more realistic values. Thus we have added thermal boundary impedance, as well as considering that the highly stressed part of the diamond (culet region) will have a reduced conductivity due to deformation and defects, etc. Another interesting aspect was the thermal conductivity of the liquid molecular hydrogen that originally was thought to be much smaller than that of the LMH.  To estimate this we used the results of Moroe et al. \cite{Moroe2011} who studied the TC of a pressurized gas of normal H$_2$ (75\% ortho, 25\% para).  Since we expect the TC of a high-density gas to be similar to that of a liquid we examined various extrapolations to higher T and P, and surprisingly found values similar to LMH, and thus chose similar values for the FEA.

In the FEA the observation of plateaus in heating curves was a challenge and we were only able to create these using very large values of the latent heat of dissociation. Morales et al. \cite{Morales2010} have estimated the latent heat or enthalpy change/atom in their DFT and quantum Monte Carlo analysis of the PPT. At $T=1000$ K and a pressure of 140 GPa, they found $\rho L = 1500-1900$ J/cm$^3$. This corresponds to $\sim$0.04-0.05 eV/atom, whereas the dissociation energy of a free molecule is 4.47 eV/atom.  To obtain plateaus in our simulations, the latent heat needs to be a factor of about 50 larger. This corresponds to a dissociation energy of 
around $2$ eV, close to the value of 1-1.5 eV reported in shock-compression
experiments \cite{Holmes1995} for corresponding densities.

It is possible that the dissociation of molecular hydrogen in the liquid phase is a multistep process.  Norman and Saitov \cite{Norman2017_1, Norman2017_2, Norman2018} have studied the pair correlation function using DFT theory. They propose that the transition involves a two step process: molecules are ionized to form H$_2^+$; this is followed by the H$_2^+$ interacting with H$_2$ to form H$_3^+$ + H. The enthalpy of formation has not been calculated in the high density liquid phase; however we note that for a free molecule the ionization energy is $\sim 15.4$ eV \cite{Liu2009}. We also note H$_3^+$ has been observed as a stable species in ionized solid hydrogen \cite{Momose2001}. Subsequently, Pierleoni, Holzmann, and Ceperley \cite{Pierleoni2017}, who also find H$_3^+$ in their theory, have shown that it is unstable. Thus, the formation of atomic metallic hydrogen by a multistep process may be involved in the transition.

Comparing our results from Fig.~\ref{SimHeatingCurves}a)
to the heating curves measured by ZSS from Fig.~\ref{SimHeatingCurves}b), it can be seen that there is a
difference in the shape of the heating curves. ZSS measure heating curves with similar 
slopes before and after the plateau, whereas in our simulations the slope after the 
plateau is smaller than that before the plateau. 
This depends on the material parameters: for example, decreasing $k_{\text{MH}}$ 
in the simulation should mean higher temperatures since the heat is more localized, 
while increasing $k_{\text{H}_2}$ should decrease the initial slope since the heat would diffuse out
into the molecular liquid more readily. Variation of these parameters may lead to results comparable to experiment \cite{Zaghoo2016}.

In the simulation, the maximum temperature in the DAC is determined as a function of the 
laser power entering the DAC, whereas in the experimental results, the heating curves were
reported as a scaled function of the laser power measured by a power meter. Moreover,
in the experiment the laser is defocused, so that only a fraction is absorbed in the tungsten film 
and part shines on the gasket. Hence, an undetermined transfer function is needed for a quantitative comparison of the heating curves.

Finally, a comparison is made between our results and the 
results measured by ZSS for the transmittance and reflectance. In their paper ZSS used a multilayer analysis to isolate the reflectance and transmittance of hydrogen from that of the tungsten layer.  Here, in Fig. 4 we show the absolute reflectance and transmittance of the system including the tungsten layer. There are two important experimental observations: in the plateau region the reflectance change is very small while the transmittance already has a measureable decrease, and as the LMH thickens the reflectance, $R$, saturates to a bulk value, which is about 0.5 as observed in the experiment. Our simulation supports this behavior. When saturated so that there is no transmittance, the absorption of the LMH is $1-R$, or about 0.5.

\section{Determination of the Phase Line P-T Points. \label{PhaseLine}}

In a dynamic experiment on liquid metallic deuterium, with rising pressure and temperature, Knudson et al. \cite{Knudson2015} observed very weak transmission of visible light (they measure at only one wavelength in the visible) and then a steep rise in reflectance attributed to metallization. Their sample was microns thick, compared to the DAC sample of MH which was nanometers thick; this can explain the differences in observed transmission (absorption). Knudson et al. ascribed the absorption to conduction to valence band transitions as the bandgap closes down. Their experiment evidently was not sensitive to observation of a plateau, possibly due to a paucity of data points. They used the onset of the rapid rise of the reflectance as the signature of the pressure of the transition; the temperature could not be measured and was calculated. They determined a semi-empirical phase line for the PPT. An issue with this phase line was that it had a very large isotope shift when compared to the hydrogen PPT line and by extrapolation was in strong disagreement with earlier dynamic measurements on deuterium.  

	Recently, there have been two new experiments on the PPT of deuterium. Measurements at static pressures in a DAC were used to determine a phase line using two criteria: onset of plateaus and onset of reflectance \cite{Zaghoo2018}. The line using the onset of reflectance criterion was used to compare to dynamic measurements that also use this criterion. There was a large disagreement with that of Knudson et al., while there was no conflict with earlier measurements, within the uncertainties.  Comparison to the hydrogen phase line using the same criteria provided an experimental determination of the isotope shift. Celliers et al. \cite{Celliers2018} made dynamic measurements on deuterium at the National Ignition Facility. They found a phase line that differed from that of Knudson et al. \cite{Knudson2015}, and was in much better agreement with earlier measurements, the phase line of ref. \cite{Zaghoo2018}, and theory \cite{Pierleoni2016,Mazzola2018}. Just as in \cite{Knudson2015} they presented a semi-empirical phase line, as temperatures were calculated, not measured. In their measurements they first observed absorption, then reflection, and used the latter as the criterion for the phase line. They too attributed absorption to bandgap transitions. 
	
	At this point in the study of the PPT there is now reasonable agreement between recently published experiments. There is a smaller disagreement for the isotope effect measured in static experiments \cite{Zaghoo2016,Zaghoo2017,Zaghoo2018, Ohta2015} and theory \cite{Pierleoni2016}. Yet there is disagreement as to the method of determining the P-T points of the transition: either the onset of the plateau or the abrupt rise in reflectance.  Although the plateaus seem to be most appropriate from the consideration of latent heat and our FEA (see also Fig.~\ref{TRcomp}), it would be useful to have an experimental determination.  An ideal order parameter for metallization is the DC electrical conductivity.  If electrical wires were inserted into the sample one could measure the conductivity and heating curves simultaneously to make the determination.
	
	It is also possible to make a determination with only optical measurements.
	In dynamic experiments the pressure is ramped up in time.  If in the dynamic experiments optical transmission was measured at two wavelengths (say blue and red) a determination could be made.  In the case of metallization creating free electron absorption, one would see simultaneous absorption of both wavelengths.  Alternately, in the case of bandgap closure one would first see absorption in the blue and then the red with increasing pressure.  In DAC experiments the sample is isochoric (fixed density) and the pressure is essentially constant; the temperature is pulsed up for each pressure or density. In the case of creation of free electrons due to the PPT all wavelengths would be absorbed simultaneously at the transition. In the case of bandgap closure, at a pressure sufficient to close the bandgap into the visible, one would observe absorption first in the blue, and then as pressure was increased, absorption would occur in both the blue and the red. This absorption should be approximately temperature independent.  In the DAC experiments visible/IR absorption was not observed until the pressure reached the plateau region and then all wavelengths were absorbed.  Evidently, the transition takes place before the bandgap is in the visible region.  We conclude that the onset of the plateaus is the proper criterion for the transition.

\section{Conclusion \label{Conclusion}}
We have presented a complete FEA of the liquid-liquid phase transition to LMH, also known as the PPT, using a simulation program that we have developed to include the many subtle conditions in the experimental study at static pressures and high temperatures in a DAC. We calculated heating curves for a number of conditions and were able to find plateaus. To simulate reasonable plateaus comparable to those observed in experiment it was necessary to use a substantially larger value of the latent heat than has been calculated for this many-body system.  We discussed a more complicated multi-step process that may possibly explain this discrepancy and may lead to new, as of yet unexplored physics.

Optical transmittance and reflectance were also simulated, and are in reasonable agreement with the experimental results. We found that the onset of the plateau is a good indicator for the insulator-to-metal transition. We discussed experiments to determine whether a change in optical parameters is due to bandgap closure or a transition to a metallic state.
Possible further steps in this investigation could be to expand the simulation to 
two/three dimensions, to include a more accurate optical response in the transition regime, and to include better estimates of the thermal (boundary) conductivities 
of the various materials. The current simulation scheme, which makes use of the Fresnel
equations for the pulse absorption, will be useful to analyse other high-pressure experiments 
with laser-pulse heated DACs.

\begin{acknowledgments}
We would like to thank Mohamed Zaghoo, Rachel Husband, Kaan Yay and Ori Noked for suggestions, useful advice and discussions of the results. M.H. acknowledges support from the University Research Fund (BOF) of the Antwerp University. I.F.S. thanks the NSF, grant DMR-1308641 and the DoE Stockpile Stewardship Academic Alliance Program, grant DE-NA0003346 for support of this research.
\end{acknowledgments}


\begin{thebibliography}{99}
\bibitem{Wigner1935} E. Wigner and H.B. Huntington, On the Possibility of a Metallic Modification of Hydrogen.  \textit{J. Chem. Phys.}, \textbf{3}, 764 (1935)

\bibitem{Ashcroft1968} N.W. Ashcroft, Metallic Hydrogen: A High-Temperature Superconductor? \textit{Phys. Rev. Lett.} \textbf{21}, 1748 (1968)

\bibitem{McMahon2012} J.M. McMahon, M.A. Morales, C. Pierleoni and D.M. Ceperley, The properties of hydrogen and helium under extreme conditions. \textit{Rev. Mod. Phys} \textbf{84}, 1607 (2012)

\bibitem{McMinis2015} J. McMinis, R.C.C. III, D. Lee and M.A. Morales, Molecular to Atomic Phase. Transition in Hydrogen under High Pressure.  \textit{Phys. Rev. Lett.} \textbf{114}, 105305 (2015)

\bibitem{Dias2017} R. Dias and I.F. Silvera, Observation of the Wigner-Huntington transition to metallic hydrogen \textit{Science} 355, 715-718 (2017).

\bibitem{Guillot2005} T. Guillot, The interiors of giant Planets: Models \& outstanding questions. \textit{Annu. Rev. Earth Planet. Sci.} \textbf{33}, 493–530 (2005).

\bibitem{Saumon1992} D. Saumon and G. Chabrier, Fluid hydrogen at high density: Pressure ionization. \textit{Phys Rev A} \textbf{46}, 2084-2100 (1992)

\bibitem{Scandolo2003} S. Scandolo, Liquid-liquid phase transitions in compressed hydrogen from first-principles simulations. \textit{Proc. Nat. Acad. of Sciences} \textbf{100}, 3051-3053 (2003)

\bibitem{Lorenzen2010} W. Lorenzen, B. Holst and R. Redmer, First-order liquid-liquid phase transition in dense hydrogen. \textit{Phys. Rev. B} \textbf{82}, 195107 (2010). 

\bibitem{Morales2010} M.A. Morales, C. Pierleoni, E. Schwegler and D.M. Ceperley, Evidence for a first-order liquid-liquid transition in high-pressure hydrogen from ab initio simulations. \textit{PNAS} \textbf{107}, 12799-12803 (2010)

\bibitem{Pierleoni2016} C. Pierleoni, M.A. Morales, G. Rillo, M. Holtzmann and D.M. Ceperley, Liquid-liquid phase transition in hydrogen by coupled electron-ion Monte Carlo simulations. \textit{PNAS} \textbf{113}, 18, 4953-4957 (2016)

\bibitem{Mazzola2018} G. Mazzola, R. Helled and S. Sorella, Phase Diagram of Hydrogen and a Hydrogen-Helium Mixture at Planetary Conditions by Quantum Monte Carlo Simulations. \textit{Phys. Rev. Lett.} \textbf{120}, 025701 (2018)

\bibitem{Weir1996} S. T. Weir, A. C. Mitchell, and W. J. Nellis, Metallization of Fluid Molecular Hydrogen at 140 GPa (1.4 Mbar). \textit{Phys. Rev. Lett.} \textbf{76}, 1860-1863 (1996)

\bibitem{Loubeyre2012} P. Loubeyre et al., Extended data set for the equation of state of warm dense hydrogen isotopes. \textit{Phys. Rev. B} \textbf{86}, 144115-144119 (2012).

\bibitem{Celliers2018} P.M. Celliers et al., Insulator-metal transition in dense fluid deuterium. \textit{Science} \textbf{361}, 6403, 677-682 (2018)

\bibitem{Zaghoo2016} M. Zaghoo, A. Salamat and I.F. Silvera, Evidence of a first-order phase transition to metallic hydrogen. \textit{Phys. Rev. B} \textbf{93}, 155128 (2016)

\bibitem{Zaghoo2017} M. Zaghoo, I. F. Silvera, Conductivity and Dissociation in Metallic Hydrogen with Implications for Planetary Interiors. \textit{PNAS} \textbf{114} (45), 11873-11877 (2017)

\bibitem{Knudson2015} M.D. Knudson et al., Direct observation of an abrupt insulator-to-metal transition in dense liquid deuterium. \textit{Science} \textbf{348}, 1455-1459 (2015)

\bibitem{Geballe2012} Z.M. Geballe and R. Jeanloz, Origin of temperature plateaus in laser-heated diamond anvil cell experiments. \textit{J. App. Phys.} \textbf{111}, 123518 (2012)

\bibitem{Goncharov2017} A.F. Goncharov and Z.M. Geballe, Comment on ``Evidence of a first-order phase transition to metallic hydrogen'' \textit{Phys. Rev. B} \textbf{96}, 157101 (2017)

\bibitem{Montoya2012} J.A. Montoya and A.F. Goncharov, Finite element calculations of the time dependent thermal fluxes in the laser-heated diamond anvil cell. \textit{J. App. Phys.} \textbf{111}, 112617 (2012)

\bibitem{Silvera2017} I. F. Silvera, M. Zaghoo, A. Salamat, Reply to ``Comment on `Evidence of a first-order phase transition to metallic hydrogen' ''. \textit{Phys. Rev. B} \textbf{96}, 237101 (2017)

\bibitem{Rekhi2003} S. Rekhi, J. Tempere and I. F. Silvera, Temperature determination for nanosecond pulsed laser heating. \textit{Rev. Sci. Inst.} \textbf{74}, 3820-3825 (2003).

\bibitem{LandauLifshitz} L.D. Landau and E.M. Lifshitz, \textit{Electrodynamics of Continuous Media}, p.45. Pergamon Press 1960.

\bibitem{Woodside1961} W. Woodside and J.H. Messmer, Thermal Conductivity of Porous Media. I. Unconsolidated Sands. \textit{J. App. Phys.} \textbf{32}, 1688 (1961)

\bibitem{Cappella2013} A. Cappella, J.-L. Battaglia, V. Schick, A. Kusiak, A. Lamperti, C. Wiemer and B. Hay, High temperature thermal conductivity of amorphous Al$_2$O$_3$ thin films grown by low temperature ALD. \textit{Adv. Eng. Mater.} \textbf{15}, 1046-1050 (2013). 

\bibitem{Hopkins2007} P.E. Hopkins, R.N. Salaway, R.J. Stevens and P.M. Norris, Temperature-Dependent Thermal Boundary Conductance at Al/Al$_2$O$_3$ and Pt/Al$_2$O$_3$ interfaces. \textit{Int. J. Thermophys} \textbf{28}: 947-957 (2007)

\bibitem{Rillo2019} G. Rillo, M.A. Morales, D.M. Ceperley and C. Pierleoni, Optical properties of high-pressure fluid hydrogen across molecular dissociation. \textit{PNAS} \textbf{116}, 20, 9770-9774 (2019)

\bibitem{Hanfland1994} M. Hanfland, R.J. Hemley, H.-K. Mao, in \textit{High-Pressure Science and Technology-1993 (Proceedings of the Joint International Association for Research and Advancement of High Pressure Science and Technology and Americal Physical Society Topical Group on Shock Compression of Condensed Matter Conference held at Colorado Springs, Colorado, June 28 - July 2, 1993)}, S.C. Schmidt, J.W. Shaner, G.A. Samara, M. Ross, Eds. (AIP Press, New York, 1994, pp. 877-880

\bibitem{Evans1998} W. J. Evans and I. F. Silvera, Index of refraction, polarizability, and equation of state of solid molecular hydrogen. \textit{Phys. Rev. B} \textbf{57}, 14105 (1998)

\bibitem{Moroe2011} S. Moroe et al., Measurements of Hydrogen Thermal Conductivity at High Pressure and High Temperature. \textit{Int. J. Thermophys.} \textbf{32}, 1887-1917 (2011).

\bibitem{Haynes2015} W.M. Haynes, \textit{Handbook of chemistry and physics, 95th edition}, CRC Press (2014-2015)

\bibitem{Raju1997} S. Raju, E. Mohandas and V.S. Raghunathan, The pressure derivative of bulk modulus of transition metals: An estimation using the method of model potentials and a study of the systematics. \textit{J. Phys. Chem. Solids} \textbf{58}, 9, 1367-1373 (1997)

\bibitem{Malitson1963} I.H. Malitson, Refraction and Dispersion of Synthetic Sapphire. \textit{J. Opt. Soc. Am.} \textbf{52}, 12 (1962)

\bibitem{Duan1999} W. Duan, B.B. Karki and R.M. Wentzcovitch, High-pressure elasticity of alumina studied by first principles. \textit{Am. Mineral.} \textbf{84}, 1961-1966 (1999)

\bibitem{Olson1993} J.R. Olson, R.O. Pohl, J.W. Vandersande, A. Zoltan, T.R. Anthony and W.F. Banholzer, Thermal conductivity of diamond between 170 and 1200 K and the isotope effect. \textit{Phys. Rev. B} \textbf{47}, 22 (1993)

\bibitem{Zaghoo2018} M. Zaghoo, R.J. Husband and I.F. Silvera, Striking isotope effect on the metallization phase lines of liquid hydrogen and deuterium. \textit{Phys. Rev. B} \textbf{98}, 104102 (2018)

\bibitem{Ohta2015} K. Ohta et al., Phase boundary of hot dense fluid hydrogen. \textit{Sci. Rep.} \textbf{5}, 16560 (2015)

\bibitem{Vinet1987} P. Vinet, J.R. Smith, J. Ferrante and J.H. Rose, Temperature effects on the universal equation of state of solids. \textit{Phys. Rev. B} \textbf{35}, 1945 (1987)

\bibitem{Holmes1995} N.C. Holmes, M. Ross and W.J. Nellis, Temperature measurements and dissociation of shock-compressed liquid deuterium and hydrogen. \textit{Phys. Rev. B} \textbf{52}, 15835 (1995).

\bibitem{Norman2017_1} G.E. Norman and I.M. Saitov, Ionization of Molecules at the Fluid-Fluid Phase Transition in Warm Dense Hydrogen. \textit{Doklady physics} \textbf{62}, 284-288 (2017)

\bibitem{Norman2017_2} G.E. Norman and I.M. Saitov, Critical Point and Mechanism of the Fluid-Fluid Phase Transition in Warm Dense Hydrogen. \textit{Doklady physics} \textbf{62}, 294-298 (2017)

\bibitem{Norman2018} G.E. Norman and I.M. Saitov, Plasma phase transition in warm dense hydrogen. \textit{Contrib. Plasma Phys.} \textbf{58}, 122-127 (2018) 

\bibitem{Liu2009} J. Liu et al., Determination of the ionization and dissociation energies of the hydrogen molecule. \textit{J. Chem. Phys.} \textbf{130}, 174306 (2009)

\bibitem{Momose2001} T. Momose, C. Michael Lindsay, Y. Zhang and T. Oka, Sharp Spectral Lines Observed in $\gamma$-Ray Ionized Parahydrogen Crystals. \textit{Phys. Rev. Lett} \textbf{86}, 4795 (2001)

\bibitem{Pierleoni2017} C. Pierleoni, M. Holzmann and D.M. Ceperley, Local structure in dense hydrogen at the liquid-liquid phase transition by Coupled-Electron-Ion Monte Carlo. \textit{Contrib. Plasma. Phys.} \textbf{58}, 1 (2017)

\end{thebibliography}
\end{document}